# A new analytical expression for calculation of the Coulomb potential in spherical nuclei

Hüseyin KOÇ [1], Erhan ESER [2*] and Cevad SELAM [3,4]

[1] Department of Electrical and Electronics Engineering, Faculty of Engineering, Muş Alparslan University, 49000, Muş, Turkey
[2] Department of Physics, Polatlı Faculty of Arts and Sciences, Ankara Hacı BayramVeli University, 06900, Ankara, Turkey
[3] Department of Business Administration, Faculty of Economics and Administrative Sciences, Muş Alparslan University, 49000, Muş, Turkey
[4] Azerbaijan National Academy of Sciences, Institute of Physics, Baku, Azerbaijan

*Corresponding author: eserphy@gmail.com

**Abstract**

A lot of problems of atomic and nuclear physics depend on with high accuracy to the Coulomb potential. Therefore, it is very important to carefully and accurately calculate the Coulomb potential. In this study, a new analytical expression was obtained for calculating the Coulomb potential by choosing the Fermi distribution function, which is suitable for charge distribution in nuclei. The proposed formula guarantees an accurate and simple calculation of the Coulomb potential of nuclei. Using the analytical expression obtained, the Coulomb potentials for several spherical nuclei were calculated for all values of the parameters. It is shown that the results obtained for arbitrary values of the radius are consistent with the literature data. In this study, the accepted values in literature of the two parameters (Coulomb radius $R_c$ and diffuseness parameter $a_c$) which are important for the coulomb potential are also discussed.

## 1. Introduction

As known, the solution of problems in many areas of physics (solid state physics, atomic and molecular physics, nuclear physics, astrophysics, statistical physics) depends on the calculation of the exponential type integrals such as $\int_0^\infty t^n/(e^{t-\sigma} \pm 1)\,dt$ [1-5]. For example, the calculation of the Fermi-Dirac function, which is one of the exponential type integrals and changing in [0, ∞] range, reveals the solution of many physical problems [4, 5]. At the same time, in the case of $\sigma = R_c/a_c$, the Fermi-Dirac function turns into the charge density distribution of nuclei and the nuclear the Woods-Saxon (WS) potential. This potential has an important role in atomic-molecular and nuclear physics. In the literature, the solution of these type integrals, which is changing in [0, ∞] range, is mostly achieved by reducing them to the gamma integrals [5]. However, especially in atomic-molecular and nuclear physics, there are such problems related to Coulomb potentials [proton diffusion, rapid proton capture (rp-process) processes and calculation of fermi integral in beta transitions]; integrals in these and similar problems are calculated numerically because they cannot be reduced to Gamma integrals [6].

It is well known that the WS potential, which is used to describe the interaction of a nucleon with heavy nucleus, plays an important role in microscopic physics [7]. On the other hand, the WS potential and its various modifications are very important in determining of the energy level spacing, particle number dependence of energy quantities and universal properties of electron distributions in atoms, nuclei and atomic clusters [8-11]. It has also a central role for optical potential model [11-14].

The single particle potential for a proton interacting with a spherical nucleus, which includes the nuclear, spin-orbit interactions and the Coulomb potential, can be written as follows [15- 17]:

$$V_{sp}(r) = V_N(r) + V_{ls}(r)\left(\vec{\ell}.\vec{s}\right) + V_c(r). \tag{1}$$

Where $V_N(r)$ and $V_{ls}(r)$ are the central and spin-orbit parts of the WS potential, respectively given by

$$V_N(r) = V_0^{n,p} \cdot f(r)\,, \qquad f(r) = \frac{1}{1+e^{\frac{r-R_0}{a}}}\,, \tag{2}$$

$$V_{\ell s}(r) = -\xi \frac{1}{r}\frac{dV_N(r)}{dr}\left(\vec{\ell}.\vec{s}\right). \tag{3}$$

In Eqs. (2) and (3), $V_0^{n,p}$, $R_0$, $a$ and $\xi$ parameters are defined as the depth of the potential, the radius of the nuclei, the diffuseness of nuclei and the part of spin-orbit, respectively.

The term $V_c(r)$ in Eq. (1) determines the Coulomb interaction between protons and is usually taken as Coulomb potential of a spherical nuclei with a uniform charge distribution (ucd) [6, 17-19]:

$$V_{ucd}(r) = Ze^2\begin{cases} \frac{1}{2R_c}\left(3-\left(\frac{r}{R_c}\right)^2\right), & r \le R_c \\ \frac{1}{r}\,, & r > R_c \end{cases} \tag{4}$$

$Z$ is the proton number, $e$ is positive unit charge, and $R_c$ is Coulomb radius. In general $R_0 = R_c$ is accepted. But, the Coulomb potential given by Eq. (4) is valid for cases where the nucleus is a hard sphere and the diffuseness of nucleus is zero ($a_c = 0$) (this issue will be discussed in later sections). But the nucleus is not a hard sphere and in proton rich nuclei the charge density distribution may exhibit diffuseness. The transition probability of the emitted proton depends sensitively on the long range behavior of the Coulomb potential and diffused barrier would affect the half-life of proton emitters. The most definitive information about nuclear sizes comes from electron scattering. These scattering experiments have made it clear that there is a “tail” where the density of nuclear matter decreases towards zero [6].

As it is known, the exact and sensitive calculation of the Coulomb potential will reveal the solution of many problems in nuclear physics. For example, the Coulomb potential has a major role in the determination of the energy of the Giant Dipole Resonance (GDR) [21], Magnetic Dipole Resonance (MDR) [22] or Isobar Analog Resonance (IAR) [23] and of the transition matrix elements of these resonances [24]. The Coulomb potential is also very important for the isospin prohibited beta transitions.

It is well known that $0^+ \rightarrow 0^+$ Fermi beta transitions between the states with different isospins (forbidden Fermi beta transitions) occur by the degradation of the principle of isospin conservation by the interaction of Coulomb between protons. Therefore, it is clear that the possibility of a forbidden beta transition is sensitive to the Coulomb potential of the nucleus. Thus, it is seen that the accurate and sensitive calculation of the Coulomb potential in Eq. (1) is very important.

In this study, it was presented a new analytical expression for the Coulomb potential $V_c(r)$, by selecting an appropriate function (Fermi distribution function used by Woods-Saxon) for the charge distribution in the nucleus. This approach allows an easy calculation of the Coulomb potential of the spherical nuclei. The obtained analytical expression was applied to a few spherical nuclei ($^{48}$Ca, $^{105}$Sn, $^{109}$I, $^{105}$Sn, $^{109}$I and $^{208}$Pb) and the results of the calculation were compared with the results obtained from the literature. The purpose of this study is to obtain an analytical expression for a simpler calculation of the Coulomb potential, and not to solve any physical problem. We claim that he proposed method can also be used to assess the Coulomb potential of all other spherical nuclei.

## 2. Theory

The average field potentials of the nuclei are generally explained using a potential consisting of Coulomb and nuclear parts. Through this paper we will discussed the Coulomb potential. This repulsive potential is fully determined with the assumption of a given nuclear charge distribution $\rho(r)$. The solution of the corresponding electrostatics problem gives [6, 16, 18-20]:

$$V_c(r) = e \int \frac{1}{|r-r'|} \rho_c(r') d^3 r' \qquad (5)$$

where

$$\rho_c(r') = \frac{\rho_0}{1+\exp[(r'-R_c)/a_c]} \tag{6}$$

$$\rho_0 = \frac{Ze}{\frac{4\pi}{3}R_c^3} \tag{7}$$

$$d^3r' = r'^2 \sin\theta'\, dr' d\theta' d\varphi'. \tag{8}$$

At the end, we obtain

$$\begin{aligned} V_c(r) &= e\int_0^{\infty}\int_0^{\pi}\int_0^{2\pi} \frac{1}{|r-r'|}\rho_c(r')r'^2 \sin\theta'\, dr' d\theta' d\varphi' \\ &= 4\pi e\int_0^{\infty} \frac{r'^2}{|r-r'|}\rho_c(r')dr' \end{aligned} \tag{9}$$

Expanding the function $\frac{1}{|r-r'|}$ in spherical harmonics and the spherical feature is considered, the following expression is obtained [25]:

$$\frac{1}{|r-r'|} = \begin{cases} \frac{1}{r'}, & r' \geq r \\ \frac{1}{r}, & r' < r \end{cases} \tag{10}$$

As can be seen above, when substituting Eq. (10) into Eq. (9), one is able to obtain the following expression [6, 18]:

$$V_c(r) = \frac{4\pi}{r}\int_0^r r'^2 \rho_c(r')dr' + 4\pi\int_r^{\infty} r'\rho_c(r')dr' \tag{11}$$

Substituting Eq. (6) into Eq. (11) taking into account the status of $r \leq R_c$ and $r > R_c$, the Eq.(11) takes the following form:

$$\begin{aligned} I_1(r;R_c) &= \frac{4\pi e}{r}\int_0^r r'^2 \rho_c(r')dr' \\ &= \frac{4\pi e}{r}\rho_0 \begin{cases} \int_0^r \frac{r'^2}{1+exp\left[(r'-R_c)/a_c\right]}dr', & for \quad r \leq R_c \\ \int_0^{R_c} \frac{r'^2}{1+exp\left[(r'-R_c)/a_c\right]}dr' + \int_{R_c}^r \frac{r'^2}{1+exp\left[(r'-R_c)/a_c\right]}dr', & for \quad r > R_c \end{cases} \end{aligned} \tag{12}$$

$$I_2(r;R_c) = 4\pi e\int_r^\infty r'\rho_c(r')\,dr'$$

$$= 4\pi e\rho_0 \begin{cases} \int_r^{R_c} \frac{r'}{1+exp\left[(r'-R_c)/a_c\right]}dr' + \int_{R_c}^\infty \frac{r'}{1+exp\left[(r'-R_c)/a_c\right]}dr' \;, & for \quad r \le R_c \\ \int_r^\infty \frac{r'}{1+exp\left[(r'-R_c)/a_c\right]}dr' \;, & for \quad r > R_c . \end{cases} \tag{13}$$

Thus, the Coulomb potential will consist of two terms:

$$V_c(r) = I_1(r;R_c) + I_2(r;R_c) \tag{14}$$

Eqs. (12) and (13) can be expressed by the binomial expansion theorem (for an arbitrary $n$ real value) defined by [26, 27]:

$$(x \pm y)^n = \sum_{m=0}^\infty (\pm 1)^m \; f_m(n)\, x^{n-m} y^m \tag{15}$$

where $f_m(n)$ is the binomial coefficients [3], which is given by

$$f_m(n) = \begin{cases} \frac{n(n-1)\ldots(n-m+1)}{m!}, & for \;\; integer \;\; n \\ \frac{(-1)^m\,\Gamma(m-n)}{m!\,\Gamma(-n)}, & for \;\; noninteger \;\; n \end{cases} \tag{16}$$

The expansion theorem for the function $\left(1+\exp\left[(r'-R_c)/a_c\right]\right)^{-1}$ contained in formulas (12) and (13) is as follows:

$$\frac{1}{1+\exp\left[(r'-R_c)/a_c\right]} = \begin{cases} 1+\sum_{m=1}^\infty f_m(-1)\, e^{m(r'-R_c)/a_c} \;, & for \quad r' < R_c \\ \sum_{m=0}^\infty f_m(-1)\, e^{-(m+1)(r'-R_c)/a_c}, & for \quad r' > R_c . \end{cases} \tag{17}$$

Eqs. (12) and (13) can be written the follows as:

$$I_1\left(r\,;R_c\right)=\frac{4\pi e}{r}\rho_0 \times\begin{cases}\int\limits_0^r r'^2\left(1+\sum\limits_{m=1}^{\infty} f_m\left(-1\right)e^{m\left(r'-R_c\right)/a_c}\right)dr', & for \quad r\le R_c\\ \int\limits_0^{R_c} r'^2\left(1+\sum\limits_{m=1}^{\infty} f_m\left(-1\right)e^{m\left(r'-R_c\right)/a_c}\right)dr'+\sum\limits_{m=1}^{\infty} f_{m-1}\left(-1\right)\int\limits_{R_c}^{r} r'^2\, e^{-m\left(r'-R_c\right)/a_c}dr', & for \quad r> R_c\end{cases} \tag{18}$$

$$I_2\left(r\,;R_c\right)=4\pi e\rho_0 \times\begin{cases}\int\limits_r^{R_c} r'\left(1+\sum\limits_{m=1}^{\infty} f_m\left(-1\right)e^{m\left(r'-R_c\right)/a_c}\right)dr'+\sum\limits_{m=1}^{\infty} f_{m-1}\left(-1\right)\int\limits_{R_c}^{\infty} r'\, e^{-m\left(r'-R_c\right)/a_c}dr'\ , & for \quad r\le R_c\\ \sum\limits_{m=1}^{\infty} f_{m-1}\left(-1\right)\int\limits_{r}^{\infty} r'\, e^{-m\left(r'-R_c\right)/a_c}dr'\ , & for \quad r> R_c\end{cases} \tag{19}$$

In this case, the integrals in the formulas (18) and (19) are in solvable form and finally the following analytical expressions are obtained:

$$I_1\left(r\,;R_c\right)=\frac{4\pi e}{r}\rho_0\begin{cases}\frac{r^3}{3}+\sum\limits_{m=1}^{\infty} f_m\left(-1\right)g_m\left(b,r,R_c\right), & for \quad r\le R_c\\ \frac{R_c^{\ 3}}{3}+\sum\limits_{m=1}^{\infty}\left\{f_m\left(-1\right)h'_m\left(b,R_c\right)+f_{m-1}\left(-1\right)g'_m\left(b,r,R_c\right)\right\}, & for \quad r> R_c\end{cases} \tag{20}$$

$$I_2\left(r\,;R_c\right)=4\pi e\rho_0\begin{cases}\frac{R_c^{\ 2}-r^2}{2}+\sum\limits_{m=1}^{\infty}\left\{f_m\left(-1\right)h_m\left(b,R_c\right)+f_{m-1}\left(-1\right)k_m\left(b,R_c\right)\right\}\ , & for \quad r\le R_c\\ \sum\limits_{m=1}^{\infty} f_{m-1}\left(-1\right)k'_m\left(b,r,R_c\right), & for \quad r> R_c.\end{cases} \tag{21}$$

Substituting Eqs. (7), (20) and (21) into (14), we obtain the following analytical expression for the Coulomb potential:

$$V_c(r) = Ze^2 \begin{cases} \dfrac{1}{2R_c}\left[3-\left(\dfrac{r}{R_c}\right)^2 + Q_m(r;R_c,b)\right], & for \quad r \le R_c \\ \\ \dfrac{1}{r} + Q'_m(r;R_c,b)\ , & for \quad r > R_c . \end{cases} \tag{22}$$

where

$$Q_m(r;R_c,b) = \frac{6}{R_c^{\ 2}} \lim_{L\to\infty} \sum_{m=1}^{L} \left\{ f_m(-1)\left[\frac{1}{r} g_m(b,r,R_c) + h_m(b,R_c)\right] + f_{m-1}(-1) k_m(b,R_c) \right\} \tag{23}$$

$$Q'_m(r;R_c,b) = \frac{3}{R_c^{\ 3}} \lim_{L\to\infty} \sum_{m=1}^{L} \left\{ f_m(-1)\frac{1}{r} h'_m(b,R_c) + f_{m-1}(-1)\left[\frac{1}{r} g'_m(b,r,R_c) + k'_m(b,r,R_c)\right]\right\} \tag{24}$$

and, the values of coefficients as follows:

$$b = \frac{a_c}{m} \tag{25}$$

$$g_m(b,r,R_c) = \left(2b^3 - 2b^2 r + br^2\right)e^{\frac{r-R_c}{b}} - 2b^3 e^{-\frac{R_c}{b}}\ , \tag{26}$$

$$g'_m(b,r,R_c) = 2b^3 + 2b^2 R_c + bR_c^{\ 2} - \left(2b^3 + 2b^2 r + br^2\right)e^{\frac{R_c-r}{b}}\ , \tag{27}$$

$$h_m(b,R_c) = b(R_c - b) - b(r-b)e^{\frac{r-R_c}{b}}\ , \tag{28}$$

$$h'_m(b,R_c) = 2b^3 - 2b^2 R_c + bR_c^{\ 2} - 2b^3 e^{-\frac{R_c}{b}}\ . \tag{29}$$

$$k_m(b,R_c) = b(b+R_c)\ , \tag{30}$$

$$k'_m(b,r,R_c) = b(b+r)e^{\frac{R_c-r}{b}}\ . \tag{31}$$

**3. Results and Discussions**

The Eq. (22) obtained for the Coulomb potential appears to be similar to the approximate expression given by Eq. (4). But, there are the correction terms in Eq. (22):

$$\Delta V_c = V_c - V_{ucd} = \begin{cases} \frac{Ze^2}{2R_c} Q(r; R_c, b), & r \leq R_c \\ Ze^2 Q'(r; R_c, b), & r > R_c \end{cases} \tag{32}$$

As can be seen, the Eq.(22) is transformed into the Eq.(4) at $a_c \rightarrow 0$ limit. In other words, the obtained analytical result (22) indicates that equation (4) is valid for situations where the diffuseness parameter is zero. This is evident from figure 1. Fig. 1 shows the radius dependence of $\Delta V_c$ for different diffusion parameters. The radius dependence of the Coulomb potentials calculated with Eqs. (4) and (22) for $^{48}$Ca and $^{208}$Pb isotopes are shown in Fig. 2. Similarly, Fig. 3 and Fig. 4 show the radius dependence of the Coulomb potentials for $^{105}$Sb, $^{109}$I, $^{147}$Tm and $^{156}$Ta nuclei. The right upper part of the Figures 2 - 4 also shows the radius dependence of the difference of the Coulomb potentials (ΔVc) calculated by formulas (4) and (22).

As seen from the figure, the $\Delta V_c - r$ differences for both isotopes (for both the light $^{48}$Ca isotope and the $^{208}$Pb heavy isotope) are the same. That is, the correction terms in formula (22) are identical for the isotopes in question. In other words, the correction terms do not depend on A or Z and depend only on the diffusion parameter $a_c$ (see Figure 1 . For the isotopes discussed, the correction values in the formula (22) are the same because the same value was taken for the diffusion parameter in the calculations. (note that in numerical calculations was taken $a_c = 0{,}7$ fm based on the universal parametrization). However, the experiments show that the parameters for different nuclei are different [28]. We think that it would be useful to take this issue into account when making the average field potential parameterizations.

In the figures ($\Delta V_c - r$), the dashed lines represent the position of the nuclei radius. The nuclei radius and Coulomb radius were taken equal ($R = R_c$) in the calculations. As seen from figures, the $\Delta V_c$ difference curve has a maximum in $r = r_{max}$. Such a maximum was observed in all the nuclei taken up (see figure. 3 and 4). Isn't it a natural idea to have this maximum is at $r = R_c$? Thus, the difference between the Coulomb radius and the nucleus radius $\Delta R = R_c - R$ can also be estimated from the calculation made. The evaluations made for nuclei used in this study are shown in Table 1. In the table are shown the nuclei radius calculated according to universal parameterization (second column), the $R_c = r_{max}$ value corresponding to the maximum value of $\Delta V_c - r$ curve (third column) and the difference of Coulomb and nuclei radii (last column). It is observed that this difference is around 1 fm and, as the nuclei become heavier it increases slightly.

## 4. Conclusion

In this study, a new analytical expression (Eq. (22)) is derived (presented) for the exact solution of the Coulomb potential which plays an important role in the solution of the problems related atomic-molecular and nuclear physics. The proposed analytical formula is calculated for all parameter values and compared with the literature dates. It is seen from figure 2 – figure 4 that the results of the calculation were in good agreement with the results obtained from the literature dates for the values of arbitrary radius (r).

Eq. (4) used for the Coulomb potential in literature applies to cases where the nuclei is a hard sphere and the nuclear diffuseness parameter is zero ($a_c = 0$). But in numerical calculations for all spherical nuclei, a single diffusion parameter ($a_c = 0.7$ fm) was used, following the universal parametrization. However, in this study the accuracy of this approach is discussed. Because in the calculation results of the Coulomb potential, the value of the heavy $^{208}$Pb isotope is approximately twice that of the light $^{48}$Ca isotope (see figure 2), which naturally leads to the idea that the diffuseness parameter for different nuclei is different.

In this study, we have tried to show that the nuclear diffuseness with Equation (22) must be different from zero ($a_c \neq 0$) and have different values ($a_{c\,{}^{48}_{20}Ca} \neq a_{c\,{}^{208}_{82}Pb}$) for all spherical nuclei. The obtained results are shown in figure 1. As it is seen from the results, it would be the right approach to say that each spherical nucleus must have a different diffuseness parameter. In addition, we think that the idea is not wrong to be calculated the parameter ($a_c$) depending on the parameters of any of the nuclei. For example, while a formula such as $\mathrm{R_c = R_0 A^{1/3}}$ depending on the mass number (A) of the nucleus in Coulomb radius calculations is used, we also think that diffuseness parameter can be calculated according to any parameter such as A, Z or N of nucleus [such as $a_c = a_0 \cdot \mathrm{f(A, Z)}$].

In this study, for the calculation of the Coulomb potential, the nucleus radius and Coulomb radius were also taken equal ($R = R_c$). If $a_c = 0$ in Eq. (22) is taken (this case corresponds to Eq.(4)), the Coulomb radius is overlapped by the nucleus radius. However, as it is seen from the figures of $\Delta \mathrm{V_c} - r$, there is a maximum in the case of $r = r_{max}$. Such a maximum value was observed in all the nuclei examined (figures 2 - 4). Considering these difference, the ΔR difference table was formed. As it is seen from Table 1, the $R_c - R$ difference increases from light to heavy nucleus.

Consequently, we think: (1) the nucleus radius is different from the Coulomb radius ($R \neq R_c$), (2) The $r_{max}$ value obtained as a result of the Eq. (22) of which the diffuseness parameter is different from zero, should be equal to the Coulomb radius ($r_{max} = R_c$).

## Captions of Figures and Table

**Figure 1.** The radius dependence of the difference between Eq.(22) and Eq.(4) for different $a_c$.

**Figure 2.** The radius dependence of the Coulomb potential calculated with Eqs. (4) and (22) for $^{48}$Ca and $^{208}$Pb isotopes.

**Figure 3.** The radius dependence of the Coulomb potential calculated with Eqs. (4) and (22) for $^{105}$Sn and $^{109}$I isotopes.

**Figure 4.** The radius dependence of the Coulomb potential calculated with Eqs. (4) and (22) for $^{147}$Tm and $^{156}$ Ta isotopes.

**Table 1.** Differences between the nuclei radius and the Coulomb radius for several spherical nuclei

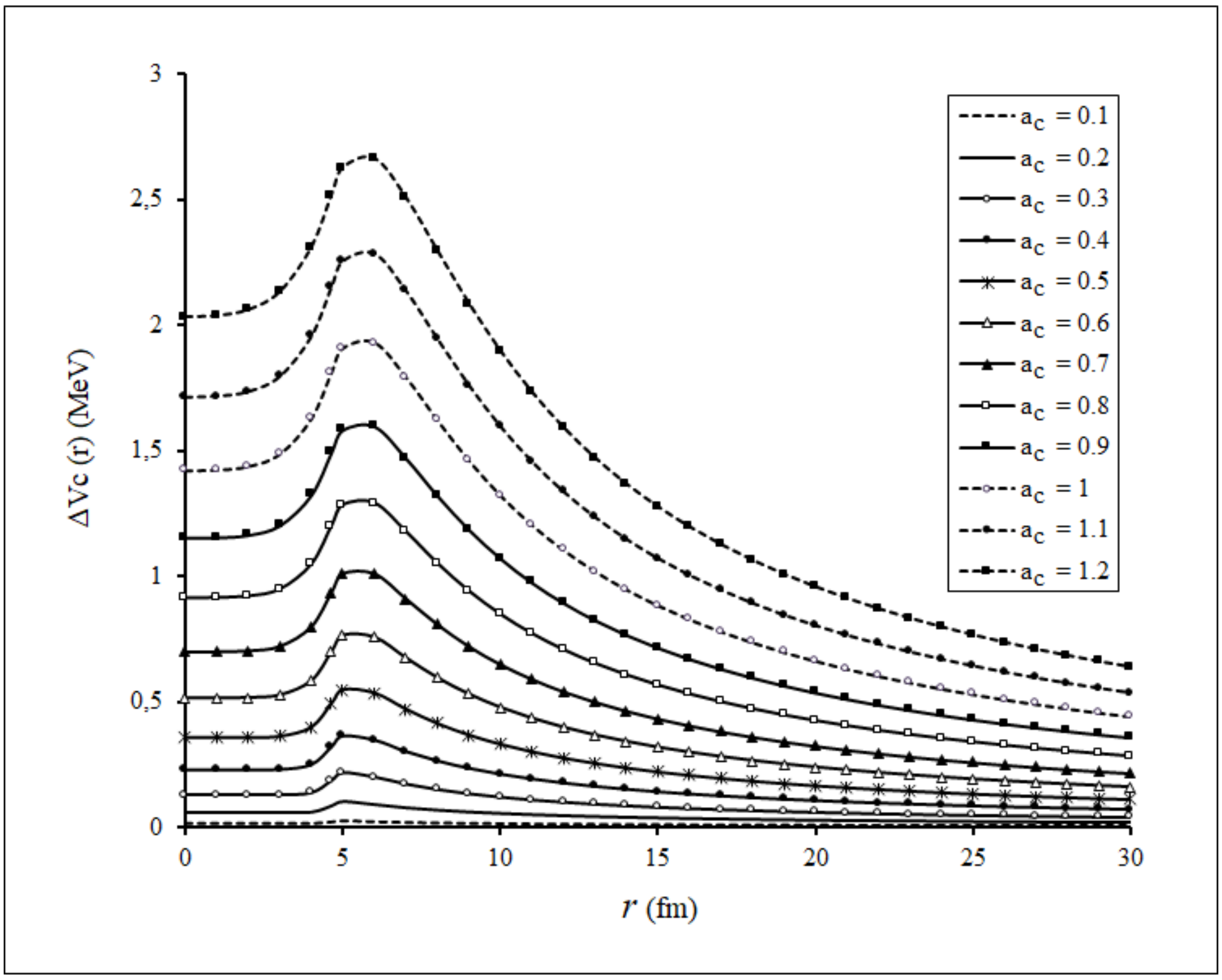

**Figure 1.**

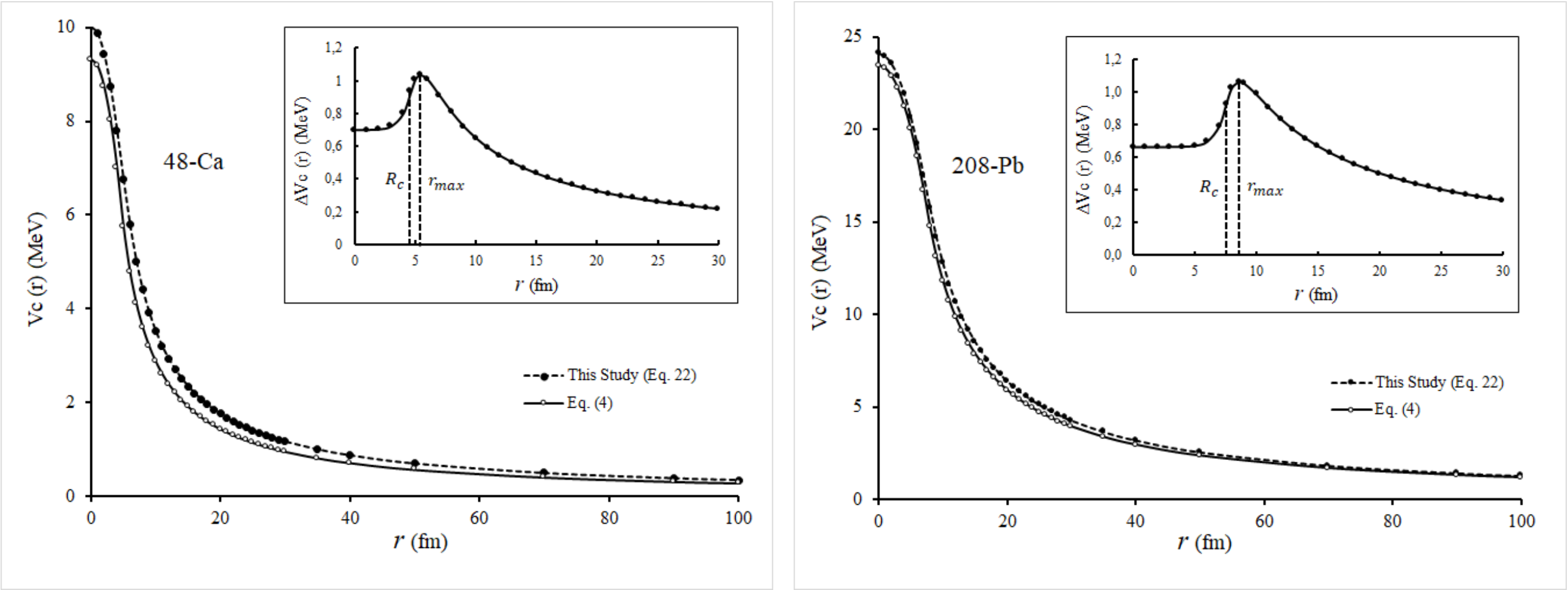

**Figure 2.**

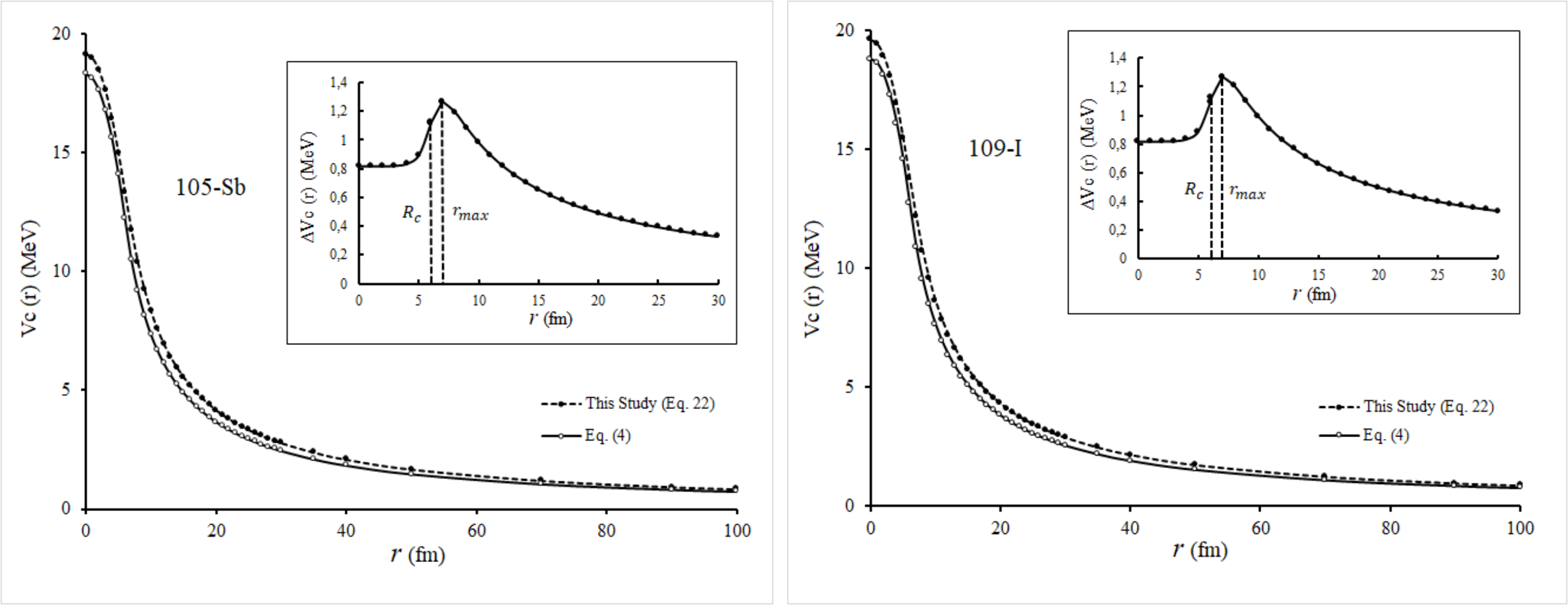

**Figure 3.**

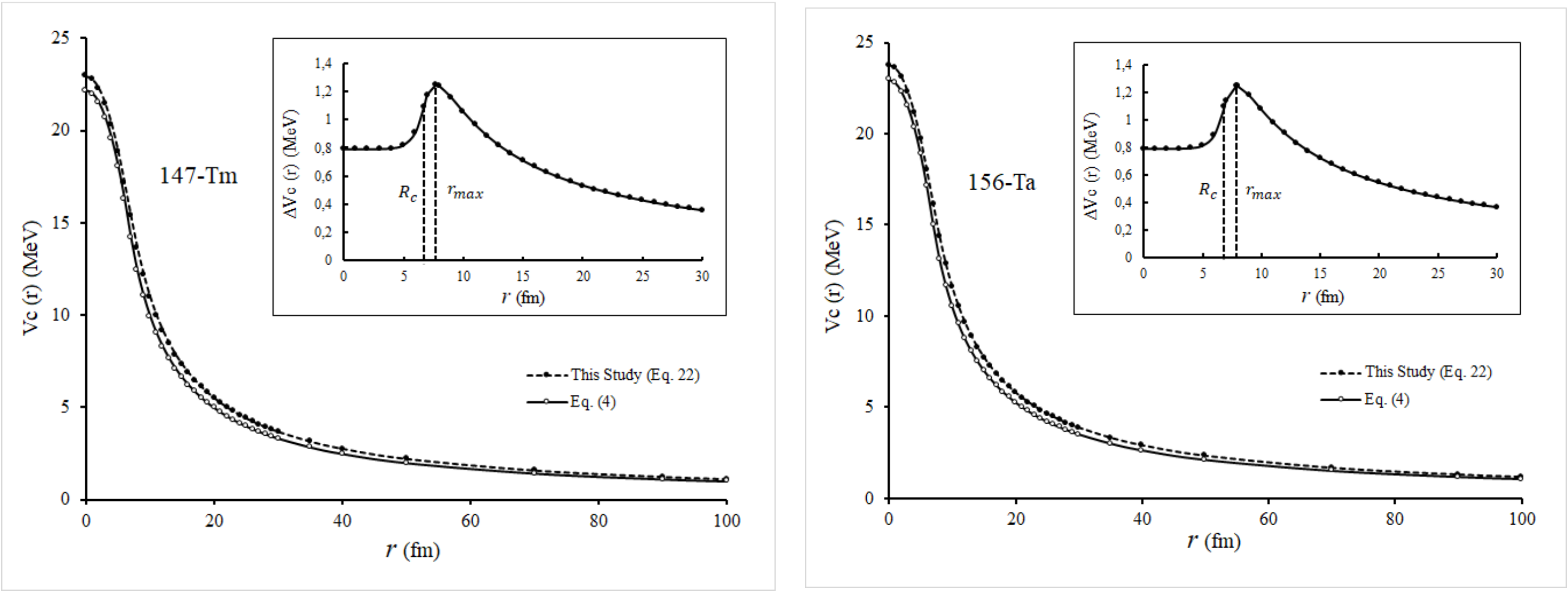

**Figure 4.**

**Table 1.**

| Isotopes | $R = r_0 A^{1/3}$ (fm) | $R_c = r_{max}$ (fm) | $\Delta R = R_c - R$ (fm) |
|---|---|---|---|
| $^{48}Ca$ | 4.63 | 5.42 | 0.79 |
| $^{105}Sb$ | 6.01 | 6.95 | 0.94 |
| $^{109}I$ | 6.09 | 7.04 | 0.95 |
| $^{147}Tm$ | 6.72 | 7.72 | 1.00 |
| $^{156}Ta$ | 6.86 | 7.87 | 1.01 |
| $^{208}Pb$ | 7.55 | 8.62 | 1.07 |